\documentstyle[preprint,aps,epsfig]{revtex}
\begin{document}
\draft
\preprint{\vbox{\hbox{IFT--P.032/98}\hbox{hep-ph/9806307}}}
\title{Testing Anomalous Higgs Couplings in Triple Photon Production at 
the Tevatron Collider}
\author{F.\ de Campos $^1$, M.\ C.\ Gonzalez--Garcia $^{2,3}$, 
        S.\ M.\ Lietti $^3$, S.\ F.\ Novaes $^3$, \\ and R.\ Rosenfeld $^3$}
\address{
$^1$ Depto.\ de F\'{\i}sica e Qu\'{\i}mica, Universidade Estadual Paulista \\
     Av.\ Dr.\ Ariberto Pereira da Cunha 333, 12500 Guaratinguet\'a, Brazil\\
$^2$ Instituto de F\'{\i}sica Corpuscular IFIC/CSIC,
     Departament de F\'{\i}sica Te\`orica\\
     Universitat de Val\`encia, 46100 Burjassot, Val\`encia, Spain\\
$^3$ Instituto de F\'{\i}sica Te\'orica, 
     Universidade  Estadual Paulista, \\  
     Rua Pamplona 145, 01405--900, S\~ao Paulo, Brazil.}
\date{May 21st}
\maketitle
\widetext
\begin{abstract}
We derive bounds on Higgs and gauge--boson anomalous interactions
using the CDF data for the process $p \bar{p} \rightarrow
\gamma\gamma\gamma + X$. We use a linearly realized  $SU_L(2)
\times U_Y(1)$ invariant effective Lagrangian to describe  the
bosonic sector of the Standard Model, keeping the fermionic
couplings unchanged. All dimension--six operators that lead to
anomalous Higgs interactions involving $\gamma$ and $Z$ are
considered.  We also show the sensitivity that can be achieved
for these couplings  at Fermilab Tevatron upgrades.
\end{abstract}
\pacs{14.80.Cp}


We certainly expect the Standard Model (SM), despite its
astonishing success in describing all the precision high energy
experimental data so far \cite{review}, to be an incomplete
picture of Nature at high energy scales. In particular, the Higgs
sector of the model, responsible for the spontaneous electroweak
symmetry breaking and for mass generation, is not fully
satisfactory since it has to be introduced in an {\it ad hoc}
fashion. Furthermore, this scalar sector has not yet been
experimentally verified.

Although we do not know the specific model which will eventually
supersede the SM, we can always parametrize its effects at low
energies by means of an effective  Lagrangian \cite{effective}
that involves operators with dimension higher than four,
containing the relevant fields at low energies and respecting the
symmetries of the Standard Model. The effective Lagrangian
approach is a model--independent way to describe new physics that
can occur at an energy scale $\Lambda$ much larger than the scale
where the experiments are performed.

The effective Lagrangian depends on the particle content at low
energies and here we will consider the possibility that the Higgs
boson can be light, being present in the higher dimensional
operators, in addition to the electroweak gauge bosons. Hence we
will use a linearly realized \cite{linear,hisz} $SU_L(2) \times
U_Y(1)$ invariant effective Lagrangian to describe the bosonic
sector of the Standard Model, keeping the fermionic couplings
unchanged. The new interactions can alter considerably the low
energy phenomenology. For instance, some operators can give rise
to anomalous $H\gamma\gamma$ and $HZ\gamma$ couplings which may
affect the Higgs boson production and decay \cite{hagiwara2}. 

It is important to notice that, since the linearly realized
effective Lagrangian relates the modifications in the Higgs
couplings to the ones in the vector boson vertex
\cite{hagiwara2,linear,hisz}, the search for Higgs bosons can be
used not only to study its properties, but also to place bounds
on the gauge boson self--interactions. This approach is more
efficient when the analysis is performed for decays of the Higgs
boson that are suppressed in the SM, such as $H \rightarrow
\gamma\gamma$ that occurs only at the one loop level, and are
enhanced by new anomalous interactions.

Events containing two photons plus additional missing energy,
photons or charged fermions represent a signature for several
models involving physics beyond the SM such as some classes of
supersymmetric  models \cite{xer}. Recently, the CDF
collaboration  have reported the search for  the signature
$\gamma\gamma + X$, where $X=$ jets, leptons, gauge bosons ($W,
\, Z, \, \gamma$) or just missing energy in $p\bar{p}$ collisions
at $\sqrt{s} = 1.8$ TeV  \cite{cdf}. Their analysis indicates a
good  agreement with the expectations from the Standard Model
(SM). In this way, they were able to set limits on the production
cross section $\sigma(p\bar{p} \to \gamma\gamma \not \!\! E_T  +
X)$ in  particular in the light gravitino scenario. 

In this work, we point out that the experimental search for
$\gamma\gamma\gamma$ events contained in the CDF analysis  can
place constraints on new physics in the bosonic sector of the SM.
For instance, associated Higgs--$\gamma$ boson production, with
the subsequent decay of the Higgs into two photons can yield this
signature. In the SM, the decay width $H \to \gamma \gamma$ is
very small since it occurs just at one--loop level \cite{h:gg}.
However, the existence of new interactions can enhance this width
in a significant way.  These anomalous Higgs boson couplings have
also been studied before in Higgs and $Z$ boson decays
\cite{hagiwara2}, in $e^+ e^-$ \cite{ee,our}, $p \bar{p}$
\cite{prl} and $\gamma\gamma$ colliders \cite{gamma}. Here we
shall show how to bound these new couplings by analyzing their
effects on the process $p \bar{p} \to \gamma \gamma \gamma + X$
at the Fermilab Tevatron.


In order to write down the most general dimension--6 effective
Lagrangian containing all SM bosonic fields, {\it i.e.\/}
$\gamma$, $W^{\pm}$, $Z^0$, and $H$, we adopt the notation of
Hagiwara {\it et al.} \cite{hisz}. This Lagrangian has eleven
independent operators in the linear representation that are
locally $SU_L(2) \times U_Y(1)$ invariant, $C$ and $P$ even. We
discard the four operators which affect the gauge boson
two--point functions at tree--level and therefore are strongly
constrained by LEP measurements. We also do not consider the
three operators that modify only the Higgs or vector boson
self--interactions,  since they  are not relevant for our
calculations. We are then left with four independent operators,
and the Lagrangian is written as,
\begin{equation}
{\cal L}_{\text{eff}} = {\cal L}_{\text{SM}} + \frac{1}{\Lambda^2} \left( 
f_{WW} {\cal O}_{WW} + f_{BB} {\cal O}_{BB} +
f_W    {\cal O}_{W} +  f_B    {\cal O}_{B}   \right) \; , 
\label{lagrangian}
\end{equation}
with the operators ${\cal O}_i$ defined as, 
\begin{eqnarray}
{\cal O}_{WW} &=& \Phi^{\dagger} \hat{W}_{\mu \nu} 
\hat{W}^{\mu \nu} \Phi \nonumber \\ 
{\cal O}_{BB} &=& \Phi^{\dagger} \hat{B}_{\mu \nu} 
\hat{B}^{\mu \nu} \Phi \\
{\cal O}_{W}  &=& (D_{\mu} \Phi)^{\dagger} 
\hat{W}^{\mu \nu} (D_{\nu} \Phi) \nonumber \\
{\cal O}_{B}  &=& (D_{\mu} \Phi)^{\dagger} 
\hat{B}^{\mu \nu} (D_{\nu} \Phi) \nonumber \; ,
\end{eqnarray}
where $\Phi$ is the Higgs field doublet, $\hat{B}_{\mu \nu} = i
(g^\prime/2) B_{\mu \nu}$, and  $\hat{W}_{\mu \nu} = i (g/2)
\sigma^a W^a_{\mu \nu}$,  with $B_{\mu \nu}$ and $ W^a_{\mu \nu}$
being the field strength tensors of the $U(1)$ and $SU(2)$ gauge
fields respectively.

Anomalous $H\gamma\gamma$, $HZ\gamma$, and $HZZ$ couplings are
generated by (\ref{lagrangian}), which, in the unitary gauge, are
given by
\begin{eqnarray}
{\cal L}_{\text{eff}}^{\text{H}} &=& 
g_{H \gamma \gamma} H A_{\mu \nu} A^{\mu \nu} + 
g^{(1)}_{H Z \gamma} A_{\mu \nu} Z^{\mu} \partial^{\nu} H + 
g^{(2)}_{H Z \gamma} H A_{\mu \nu} Z^{\mu \nu}
\nonumber \\
&+& g^{(1)}_{H Z Z} Z_{\mu \nu} Z^{\mu} \partial^{\nu} H + 
g^{(2)}_{H Z Z} H Z_{\mu \nu} Z^{\mu \nu} 
\; , 
\label{H} 
\end{eqnarray}
where $A(Z)_{\mu \nu} = \partial_\mu A(Z)_\nu - \partial_\nu
A(Z)_\mu$. The effective couplings $g_{H \gamma \gamma}$,
$g^{(1,2)}_{H Z \gamma}$, and $g^{(1,2,3)}_{H Z Z}$ are related
to the coefficients of the operators appearing in (\ref{lagrangian})
through,
\begin{eqnarray}
g_{H \gamma \gamma} &=& - \left( \frac{g M_W}{\Lambda^2} \right)
                       \frac{s^2 (f_{BB} + f_{WW})}{2} \; , 
\nonumber \\
g^{(1)}_{H Z \gamma} &=& \left( \frac{g M_W}{\Lambda^2} \right) 
                     \frac{s (f_W - f_B) }{2 c} \; ,  
\nonumber \\
g^{(2)}_{H Z \gamma} &=& \left( \frac{g M_W}{\Lambda^2} \right) 
                      \frac{s [2 s^2 f_{BB} - 2 c^2 f_{WW}]}{2 c}  \; , 
\label{g} \\ 
g^{(1)}_{H Z Z} &=& \left( \frac{g M_W}{\Lambda^2} \right) 
	              \frac{c^2 f_W + s^2 f_B}{2 c^2} \nonumber \; , \\
g^{(2)}_{H Z Z} &=& - \left( \frac{g M_W}{\Lambda^2} \right) 
  \frac{s^4 f_{BB} +c^4 f_{WW} }{2 c^2} \nonumber \; ,  
\end{eqnarray}
with $g$ being the electroweak coupling constant, and $s(c)
\equiv \sin(\cos)\theta_W$. 

The calculation of the reaction $p\bar{p}\to\gamma\gamma\gamma$
was performed using the Helas package \cite{helas}. We have
constructed new subroutines in order to incorporate the anomalous
contributions.  The irreducible background subprocesses $q
\bar{q} \to \gamma \gamma \gamma$ for $q = u, \, d, \, s$ were
generated by MadGraph \cite{mad} and the new contributions were
included. In this way, all the anomalous contributions and their
respective interference with the SM were evaluated. We convoluted
this subprocess cross sections, using a Monte Carlo integration
\cite{vegas}, with the corresponding parton distributions using
the MRS (G) \cite{pdf} set of proton structure functions with the
scale given by the parton--parton center--of--mass energy. 

The CDF Collaboration \cite{cdf} search for anomalous
$\gamma\gamma$ events has included events that have two photons in
the central region of the detector ($|\eta| < 1$), with a minimum
transverse energy of 12 GeV, plus an additional photon with $E_T
> 25$ GeV. The photons were required to be separated by an angle
larger the $15^\circ$. After applying these cuts, no event was
observed, while the expect number from the background is $0.1 \pm
0.1$ in the 85 pb$^{-1}$ collected. Therefore, at 95 \% CL this
experimental result  implies that the signal should have less
than 3 events. The efficiency of identification of an isolated
photon is $68 \pm 3 \%$, for $E_T > 12$ GeV, and grows to $84 \pm
4 \%$, for $E_T > 22$ GeV. We have taken into account these
efficiencies in our estimate. 

It is important to notice that the dimension-six operators
(\ref{lagrangian}) do not induce $4$--point anomalous couplings
like $Z Z \gamma \gamma$, $Z \gamma\gamma \gamma$, and $\gamma
\gamma \gamma \gamma$, being these terms generated only by
dimension--eight and higher operators. Since the production and
decay of the Higgs boson also involve two dimension--six
operators, we should, in principle, include in our calculations
dimension--eight operators that contribute to the above
processes. Notwithstanding, we can neglect the higher order
interactions and bound the dimension--six couplings under the
naturalness assumption that no cancelation takes place amongst
the dimension--six and --eight contributions that appear at the
same order in the expansion. 

\begin{figure}
\begin{center}
\mbox{\epsfig{file=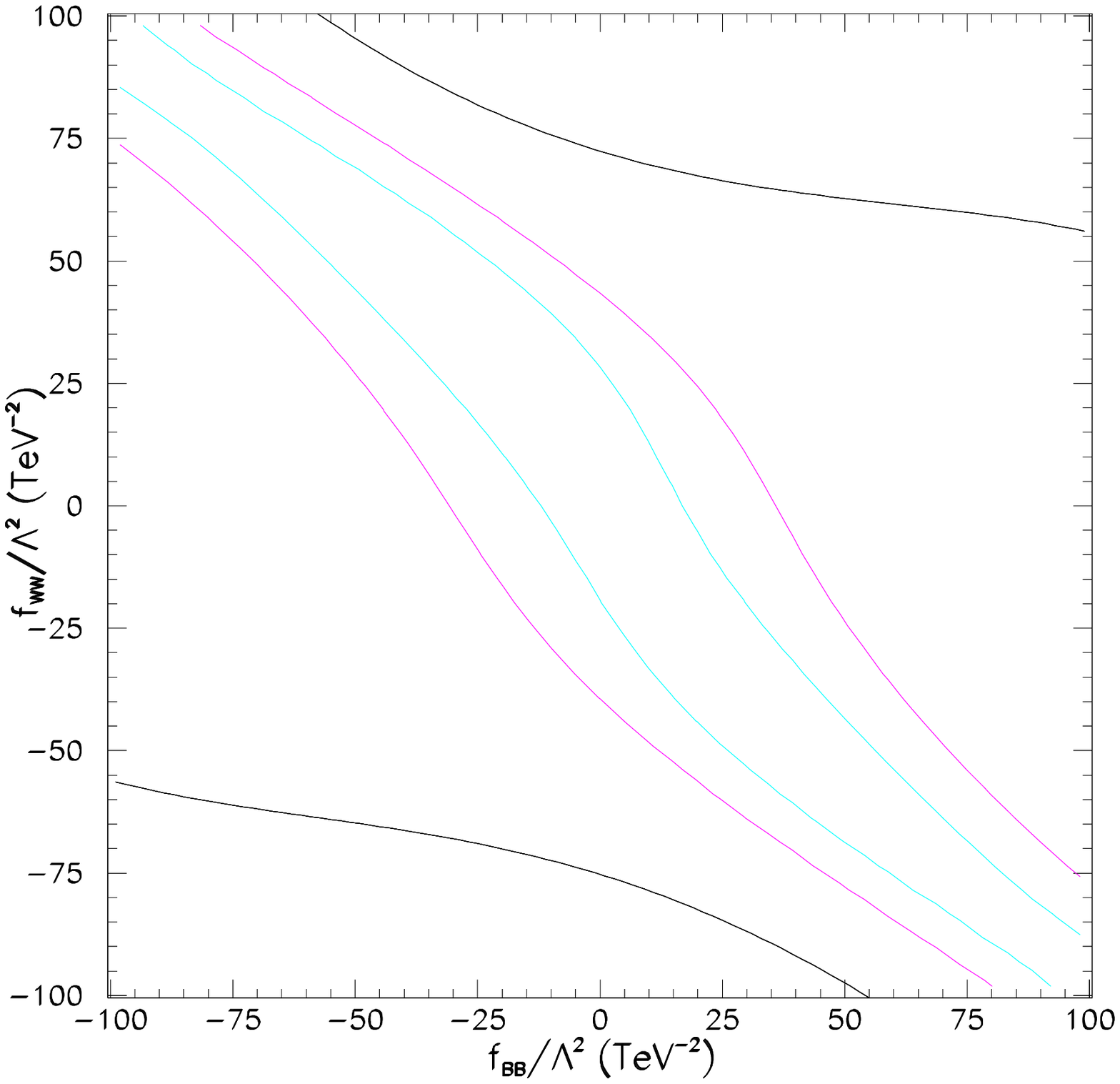,width=0.9\textwidth}}
\end{center}
\caption{Exclusion region outside the curves in the  $f_{BB}
\times f_{WW}$ plane, in TeV$^{-2}$, based on the CDF analysis
\protect\cite{cdf} and Tevatron  upgrades of $\gamma\gamma\gamma$
production, assuming $ M_H = 100$ GeV.  The curves show the 95\%
CL deviations from the SM total cross section.  The outermost
curves are based on the CDF analysis, the  intermediate curves on
the Tevatron Run II analysis, and the innermost  curves are based
on the Tevatron TeV33 upgrade.}
\label{exc:cdfupg}
\end{figure}

We start our analysis by examining which are the bounds that can
be placed on the anomalous coefficients from the negative search
of 3 photon events made by the CDF Collaboration. We start by
assuming that the only non--zero coefficients are the ones that
generate the anomalous $H\gamma\gamma$, {\it i.e.}, $f_{BB}$ and
$f_{WW}$.   Our results for the 95\% CL exclusion region in the
plane $f_{BB} \times f_{WW}$, obtained from the CDF data, are
presented in Fig.\ \ref{exc:cdfupg}. For $f_{BB} = - f_{WW}$ the
anomalous contribution to $H\gamma\gamma$ becomes zero,
independently of the values of $f_{W}$ and $f_{B}$, and the
bounds become very weak in this region. 

As mentioned above, the coupling $H\gamma\gamma$ derived in  Eq.\
(\ref{g}) involves $f_{WW}$ and $f_{BB}$. In consequence, the
anomalous $\gamma\gamma$ signature is only possible when those
couplings are non--vanishing. The couplings $f_B$ and $f_W$, on
the other hand, affect the production mechanisms for the Higgs
boson. In order to reduce the number of free parameters  one can
make the assumption that all blind operators affecting the Higgs
interactions have a common coupling $f$, {\it i.e.} $f = f_W =
f_B = f_{WW} = f_{BB} = f$ \cite{review,hisz,hagiwara2}. In this
scenario we can relate the Higgs boson anomalous coupling $f$
with the LEP conventional parametrization of the vertex $WWV$
($V=Z$, $\gamma$) \cite{lepconv} can be written as, 
\begin{equation}
\alpha = \alpha_{B \Phi} = \alpha_{W \Phi} = 
\frac{M_W^2}{2 \Lambda^2}~ f \; .
\label{trad} 
\end{equation}

Table \ref{tab:f} shows the 95\% CL allowed region of the
anomalous couplings in the above scenario. As could be expected,
these bounds become weaker as the Higgs boson mass increases. We
also show the related bounds in $\alpha = \alpha_{B \Phi} = 
\alpha_{W \Phi}$ in Table \ref{tab:alpha}.

\begin{figure}
\begin{center}
\mbox{\epsfig{file=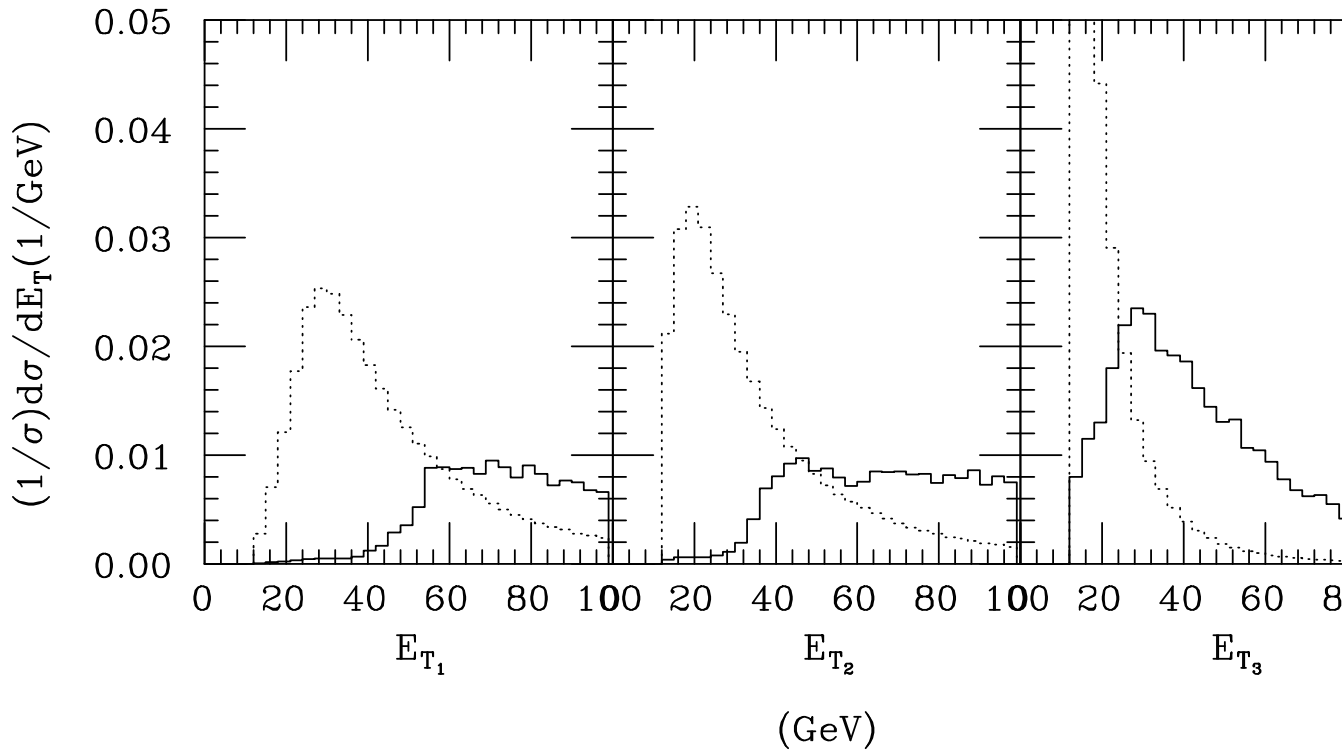,width=1.1\textwidth,angle=90}}
\end{center}
\vskip -3 cm
\caption{Transverse momentum distribution of the three photons for
$\protect\sqrt{s}= 2$ TeV, for the SM background (dotted line) and for the 
anomalous contributions (full line). We have taken $ M_H = 100$ GeV, and 
$f_i/\Lambda^2 = 100$ TeV$^{-2}$.}
\label{et:dis}
\end{figure}

We now extend our analysis to the upgraded Tevatron collider. We
first study the possible improvements in the kinematical cuts  in
order to get better sensitivity to the anomalous coefficients.
First of all, we order the three photons according to their
transverse energy, {\it i.e.\/} $E_{T_{1}} > E_{T_{2}} >
E_{T_{3}}$, and we adopt a preliminary cut of $E_{T_{i}} > 12$
GeV and $|\eta_{i}| < 1$, for all the three photons.  In Fig.\
\ref{et:dis}, we show the transverse energy distribution for the
three photons for $\sqrt{s} = 2$ TeV. Comparison is made between
the SM background and the new anomalous distribution for $f =
100$ TeV$^{-2}$, and for a Higgs boson mass of $100$ GeV. 

These distributions strongly suggest that a cut on the transverse
energy of  the most energetic photon with a simultaneous cut in
transverse energy of the two softest photons can improve the
sensitivity. We tried two sets of cuts: (a)  $E_{T_{1}} > 40$ GeV
while  $E_{T_{2,3}} > 25$ GeV, and (b)   $E_{T_{1}} > 40$ GeV,
with $E_{T_{2,3}} > 12$ GeV.  Cut (a) leads to a large background
reduction of a factor 5.5 but it also  reduces the number of
signal events by a factor two 2. So the significance  of the
signal  over the background
($S=N_{\text{Signal}}/\sqrt{N_{\text{Background}}}$)  is enhanced
only by 17\%. Cut (b) however leads to a smaller background
rejection of a factor of 2 without  significantly changing the
signal. The significance is now improved by a factor of 41\%, so
we present our results considering this set of cuts. We always
require the photons to be in the central region of the detector
($|\eta_{i}| < 1$) where there is sensitivity for electromagnetic
showering. In our estimates we assume the same detection
efficiency for photons as the present CDF efficiencies given
above.

After applying the cuts, we obtain the 95\% CL exclusion
region in the plane $f_{BB} \times f_{WW}$ shown in  Fig.\
\ref{exc:cdfupg}. We have assumed that the upgraded Tevatron
collider will reach a centre--of--mass energy of $\sqrt{s} = 2$
TeV with an integrated luminosity of 1 fb$^{-1}$, in the Run II,
and of 10 fb$^{-1}$, in the TeV33 run \cite{tevatron}. Again, the
$f_{BB} = - f_{WW}$ line is unbounded since  the anomalous
contribution to $H\gamma\gamma$ is zero in this case.

In Table \ref{tab:f}, we present the 95\% CL limit of the
anomalous couplings when all couplings are taken to be equal, for
different Higgs boson masses. The associated bounds in
$\alpha=\alpha_{B\Phi}= \alpha_{W\Phi}$ are also shown in Table
\ref{tab:alpha}.  These bounds are comparable with the
preliminary results of the combinations  of measurements from the
individual LEP and D\O~ experiments  \cite{moriond},
$\alpha_{B\Phi}=-0.05^{+0.22}_{-0.20}$, and
$\alpha_{W\Phi}=-0.03^{+0.06}_{-0.06}$. The comparison is to be
taken with  a pinch of salt as the LEP--D\O~ bounds  are given for
only one coupling different from zero while our bounds  hold for
$\alpha_{B\Phi}=\alpha_{W\Phi}$

Summarizing, in this work we have estimated the limits on
anomalous dimension--six Higgs boson interactions that can be
derived from the investigation of three photon events at the
Fermilab Tevatron.  We have used the present data from the CDF
collaboration and we have estimated the attainable sensitivity at
the upgraded Tevatron. Under the assumption of equal coefficients
for all anomalous Higgs operators, these bounds also lead to
limits on triple--gauge--boson couplings. 


\acknowledgments
We would like to thank Alexander Belyaev for very useful
discussions. This work was supported by Conselho Nacional de
Desenvolvimento Cient\'{\i}fico e Tecnol\'ogico (CNPq), and by
Funda\c{c}\~ao de Amparo \`a Pesquisa do Estado de S\~ao Paulo
(FAPESP).



\begin{table}
\begin{tabular}{||c||c||c||c||}
$M_H$(GeV) & \multicolumn{3}{c||}{$f/\Lambda^2$(TeV$^{-2}$)} \\
\hline 
\hline
  & CDF & Tevatron Run II & Tevatron TeV33 \\
\hline 
\hline
100 & ( $-$61.7 , 64.5 ) & ( $-$23.2 , 23.3 ) & ( $-$13.7 , 13.9 ) \\
\hline
120 & ( $-$75.5 , 76.9 ) & ( $-$25.0 , 25.0 ) & ( $-$14.4 , 14.5 ) \\
\hline
140 & ( $-$92.0 , 93.2 ) & ( $-$29.1 , 29.5 ) & ( $-$15.3 , 15.7 ) \\
\hline
160 & ( $-$113 , 115 ) & ( $-$34.0 , 35.8 ) & ( $-$16.1 , 17.8 ) 
\end{tabular}
\medskip
\caption{The minimum and maximum values (min, max) of
$f/\Lambda^2$, at 95\% CL, from  
$\gamma\gamma\gamma$ production at CDF and Tevatron upgrades, 
assuming that all $f_i$ are equal.}
\label{tab:f}
\end{table}

\begin{table}
\begin{tabular}{||c||c||c||c||}
$M_H$(GeV) & \multicolumn{3}{c||}{$\alpha=\alpha_{B \Phi}=\alpha_{W \Phi}$} \\
\hline 
\hline
  & CDF & Tevatron Run II & Tevatron TeV33 \\
\hline 
\hline
100 & ( $-$0.197 , 0.206 ) & ( $-$0.074 , 0.075 ) & ( $-$0.044 , 0.044 ) \\
\hline
120 & ( $-$0.242 , 0.246 ) & ( $-$0.080 , 0.080 ) & ( $-$0.046 , 0.046 ) \\
\hline
140 & ( $-$0.294 , 0.298 ) & ( $-$0.093 , 0.094 ) & ( $-$0.049 , 0.050 ) \\
\hline
160 & ( $-$0.362 , 0.368 ) & ( $-$0.109 , 0.115 ) & ( $-$0.052 , 0.057 ) 
\end{tabular}
\medskip
\caption{The minimum and maximum values (min, max) of
$\alpha = \alpha_{B \Phi}=\alpha_{W \Phi}$, at 95\% CL, from  
$\gamma\gamma\gamma$ production at CDF and Tevatron upgrades, 
assuming that all $f_i$ are equal.}
\label{tab:alpha}
\end{table}

\end{document}